\documentclass{./acm_proc_article-sp}
\usepackage{epic,eepic,epsf}
\usepackage{subfigure}
\usepackage{cite}
\usepackage[ruled]{algorithm}
\usepackage{algorithmic}
\usepackage{color}
\usepackage[table]{xcolor}
\usepackage{multirow}

\begin{document}

\conferenceinfo{ }{ }
\title{An Autonomous Distributed Admission Control Scheme for IEEE 802.11 DCF}
\numberofauthors{2}
\author{
\alignauthor
Preetam Patil\\
\affaddr{Department of CSE}\\
\affaddr{Indian Institute of Technology - Bombay, India}\\
\email{pvpatil@iitb.ac.in}
\alignauthor
Varsha Apte\\
\affaddr{Department of CSE}\\
\affaddr{Indian Institute of Technology - Bombay, India}\\
\email{varsha@cse.iitb.ac.in}
}

\maketitle
\begin{abstract}
Admission control as a mechanism for providing QoS requires an accurate description
of the requested flow as well as already admitted flows. Since 802.11 WLAN capacity
is shared between flows belonging to all stations, admission control requires
knowledge of all flows in the WLAN. Further, estimation of the 
load-dependent WLAN capacity through analytical model requires inputs about 
channel data rate, payload size and the number of
stations. These factors combined point to a centralized admission control
whereas for 802.11 DCF it is ideally performed in a distributed manner.
The use of measurements from the channel avoids explicit inputs about the
state of the channel described above.
BUFFET, a model based measurement-assisted distributed admission control 
scheme for DCF proposed in this paper relies on measurements to derive
model inputs and predict WLAN saturation, thereby maintaining average 
delay within acceptable limits.
Being measurement based, it adapts to a combination of data rates and 
payload sizes, making it completely autonomous and distributed.
Performance analysis using OPNET simulations suggests that BUFFET is able to
ensure average delay under 7$ms$ at a near-optimal throughput. 
\end{abstract}
\newpage
\section{Introduction}
With the widespread use of WLANs based on IEEE 802.11 distributed
coordination function (DCF),
efforts are on to improve the Quality of Service (QoS) offered by WLANs.
The most important component of QoS is the delay experienced by 
packets. While real-time flows have strict requirements
on delays, all applications remain sensitive to high and variable delays.

The proposed QoS-oriented 802.11e standard provides prioritized access 
through the Enhanced Distributed Channel Access (EDCA),
but 802.11e devices are not widely available.
Moreover, the QoS provision of 802.11e EDCA depends on appropriate
configuration of the tunable parameters and admission control,
otherwise its performance degrades to that of DCF.
On the other hand, it has been argued \cite{zhai_how} that
DCF is capable of providing acceptable delays as long as the load
on WLAN is maintained within the capacity of the WLAN.

Provision of QoS in communication systems necessarily involves 
maintaining the load within the finite capacity of the system. 
This task is performed
by the call \footnote{the term `call' is used synonymously with `flow' 
in this context.} 
admission control (CAC) mechanism based on a selected criterion. Admission
control can be performed on a per-flow or per-host basis,
either before admitting an entity, or, in some cases, even after admission,
if it is clear that the desired (or guaranteed) QoS requirements can not
be met.

Various models proposed
for DCF and EDCA help predict the achievable throughput and delay
\cite{bianchi:model, engelstad:delay}.
However, the application of these models for admission control requires an exact 
description of the traffic parameters such as packet arrival rate,
average packet size, as well as WLAN parameters.
Our previous experience \cite{patil:sizing} suggests that describing
the packet stream at the link layer is difficult 
due to the diverse application characteristics as well
as control overheads of the intermediate layers. 

In addition, since the WLAN capacity is shared between all stations,
the capacity computation
requires the statistics of all flows in the WLAN that in turn lends itself to a
centralized mode of admission control. 

In order to preserve the advantages of the distributed operation of DCF,
our endeavor is to design an autonomous, distributed admission control
that requires minimal inputs and is able to deduce current state of the
WLAN from relevant measurements. The use of measurements from the channel 
or the WLAN interface will serve two purposes:
\begin{list}{$\bullet$} {\setlength{\itemsep}{0.01in}
  \setlength{\topsep}{0.02in} \setlength{\leftmargin}{0.25in}}
  \item help the station estimate the aggregate statistics for
    its admitted flows that are difficult to characterize.
  \item help the station deduce traffic statistics for other stations' flows
    without using any message passing.
\end{list}

In this paper, we present one such
distributed admission control scheme named BUFFET. In the next section 
we present a summary of related work that motivates the need for current
work. Section \ref{ac_80211} presents
the admission control problem and the analytical framework for the solution.
Section \ref{algo_meamo} describes the algorithm in detail as well as the
two competing approaches we use for performance comparison.
Performance analysis of BUFFET and other approaches is presented
in Section \ref{perf}. 
We conclude with a discussion of performance results
and future work in Section \ref{conclusion}.

\section{Related Work}\label{related}

Bianchi and Tinnirello \cite{bianchi:kalman} use the
collision probability $p$ derived from the measured transmission 
probability to estimate the number $n$ of competing stations. 
%
Pong and Moors \cite{pong:call} propose a call admission
control for EDCA based on the saturation model in \cite{bianchi:model}.
Per-flow measurement of the collision probability $p$ is used to 
estimate the throughput achievable by the flow. 
A limitation of saturation model based CAC is that the model exaggerates
the effect of contention, especially at higher $n$.

The centralized CAC for EDCA
proposed by Kong et al. \cite{kong:measurement} uses the measured channel
utilization to estimate 
the achievable bandwidth (fair share of the throughput) for the requested flow
based on a non-saturation model. 
%
The CAC for EDCA proposed by Bai et al. \cite{bai:admission} 
attempts to keep the queue utilization ($\rho$) below a threshold. 
$\rho$ is computed using regression analysis and an analytical model from
the measured value of $\tau$ (the transmission probability by a station 
in a slot) and load specification. 
%

An important shortcoming of the CAC mechanisms listed above is that they 
require exact specification of packet arrival rates (except
saturation model based CACs) and payload size \emph{for all flows}.
It implies a centralized CAC mechanism that uses and stores this 
information for admission decision.

It is possible that a flow obtains more than its
fair share of bandwidth (WLAN throughput/$n$) without violating 
the QoS of other flows as long as the WLAN is not saturated. A uniform
throughput sharing assumption \cite{pong:call,kong:measurement}
results in rejecting such flows, even if
there is spare capacity.

Channel utilization (fraction of channel time used by transmissions)
threshold based CAC has been explored in
\cite{chou_achieving, zhai:admission, gu:measurement, bazzi:measurement}.
The CAC proposed by Chou et al. \cite{chou_achieving} 
maintains the allocated airtime below a threshold, but the
airtime computation excludes the overheads of contention mechanism.
Admission control for DCF proposed in \cite{zhai:admission}
combines channel utilization (including the requested flow) threshold based
CAC for real-time traffic and rate control for best-effort traffic.
The CAC scheme for EDCA in \cite{gu:measurement} uses the measured
utilization to decide on accepting a new flow or stopping low-priority
flows. The work in \cite{bazzi:measurement} evaluates two threshold-based
schemes for infrastructure WLANs, based on
channel utilization and access point queue size respectively.

Performance of threshold based CAC schemes is contingent on
the selection of the correct threshold especially in realistic WLAN scenarios
because the optimum value of the threshold depends
on payload, channel data rate, and number of stations.

\section{Admission Control in 802.11 WLANs} \label{ac_80211}
The link capacity of a 802.11 WLAN varies depending
on traffic parameters such as the number of stations, packet size, and channel
data rate \cite{zhai_how}.
The admission control for WLANs is further complicated
by the requirement of predicting the capacity or the delay at the 802.11 MAC.
The admission control objective in this paper is to keep 
the average delay for all flows within acceptable limits. 
Thus the admission control provides statistical QoS guarantees.

\subsection{Requirements and desired properties of distributed CAC}
We start with the following design objectives for the distributed CAC mechanism:
\begin{list}{$\bullet$} {\setlength{\itemsep}{0.01in}
  \setlength{\topsep}{0.02in} \setlength{\leftmargin}{0.25in}}
  \item The algorithm is to run at every wireless station without
    requiring any centralized control and coordination.
  \item No knowledge of global parameters would be assumed; 
    measurements are made locally at the WLAN interface.
  \item The measurements or the algorithm will not necessitate any change to the 
    802.11 protocol.
\end{list}
The following are the desired properties of a measurement-based 
admission control algorithm:
\begin{list}{$\bullet$} {\setlength{\itemsep}{0.01in}
  \setlength{\topsep}{0.02in} \setlength{\leftmargin}{0.25in}}
  \item The algorithm should be responsive to changing load and number of
    stations.
  \item It should adapt to varying data rates
    selected by stations based on channel quality.
  \item It should not depend on accurate statistical characteristics
    of all flows.
  \item It should be scalable with respect to the number of flows as 
    well as stations.
\end{list}

\subsection{Using analytical model of 802.11 MAC to predict saturation}
\begin{figure}[htbp]
  \begin{center}
    	\includegraphics[width=3.0in]{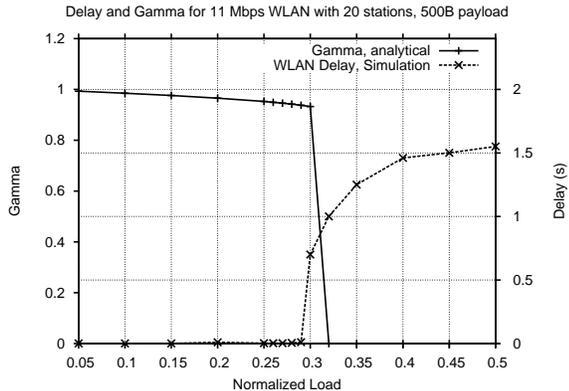}
  \end{center}
  \caption{$\gamma$ and delay co-relation}
  \label{fig:gamma_delay}
\end{figure}

The delay experienced by a packet is the queueing delay at WLAN 
interface plus the time to transmit the packet (including contention and 
collisions, if any). This queue can be modeled as an $M/G/1$ queue assuming
Poisson arrival process. The service rate of the queue is however dependent 
on the arrival rate $\lambda$. 
As illustrated by the delay vs. load curve in Fig. \ref{fig:gamma_delay}, 
the delay is close to nominal packet transmission time at moderate loads
whereas it increases by an order of magnitude after the WLAN saturates
(e.g., at 29\% load in Fig. \ref{fig:gamma_delay}).
Thus prevention of WLAN saturation has the desired effect of
maintaining average delay within acceptable limits.

Avoiding saturation requires  predicting it in real time
given the state of current load, requested load, and WLAN parameters.
For this purpose, we use a Discrete Time Markov Chain (DTMC) based 
analytical model we have developed. It is an extended version of 
the model in \cite{bianchi:model} and is
applicable to saturated as well as non-saturated WLANs.
We add a state called \emph{idle} to the single-station DTMC in \cite{bianchi:model}.
A station, after completing a post-backoff (mandatory backoff after a
successful transmission), examines the state of the interface queue.
We define a probability $\gamma$ as the probability of queue being empty
with which the station enters \emph{idle} state after post-backoff. 
The estimated $\gamma$ after accounting for an incoming flow can be used
as an indicator of saturation (as illustrated in Fig. \ref{fig:gamma_delay}).
$\gamma$ can be obtained from the model as a function 
$\Psi(\lambda,n, T_s)$ (equation \eqref{eqn:gamma})
of three variables $\lambda$ (packet arrival rate),
$n$ and $T_s$ (duration of successful transmission).
The details of the DTMC and computation of $\gamma$ are presented in the 
Appendix.
Based on this argument, the CAC algorithm follows.

\section{Model based Distributed Admission Control (BUFFET)}
\label{algo_meamo}
In a distributed CAC scheme, a station may not be aware of the flows 
belonging to other stations, neither will it know the instantaneous data rates 
adopted by individual stations. However, the station is capable of
listening to all transmissions and derive inference about the current
load on the WLAN as well as WLAN parameters. 
The load/population dependent inputs to the model are generated
by combining the measurements with the flow specifications provided by 
the application for the requested flow.
The expected value of $\gamma$ computed using the model is used to
decide on accepting/rejecting the flow.
The point of departure for BUFFET as compared to
other techniques is that no external inputs other than
the flow specification for the requested flow are required.

\subsection{WLAN interface measurements}
We follow the convention of denoting the measurement samples by $\hat{ }$
(e.g., $\hat{T}$) and running averages by $\bar{ }$ (e.g., $\bar{T}$).

\subsubsection{Frame transmission rate $\hat{R}_{tx}$}
As the station has no means of measuring the packet arrival rate
at other stations, we derive the aggregate packet arrival rate to the WLAN,
$\lambda_{MAC}$ from the measured rate of packet 
transmissions (successful and collisions) $\bar{R}_{tx}$.

\subsubsection{Average transmission slot duration $\hat{T}_{tx}$}
The throughput of a non-saturated WLAN is greatly influenced
by the average duration of a frame transmission which in turn depends on
the average frame size for all frames (including higher layer control 
frames) and PHY data rates used by the transmitting station.

A radio interface is capable of measuring the average duration
$\hat{T}_{tx}$ of transmission. This single measurement abstracts
out the effect of the two important variable parameters mentioned above
and it suffices because the model requires just the duration of successful
and collided transmissions ($T_s$ and $T_c$ respectively).

\subsubsection{The number of stations with active flows, $n$}
The number of active stations ($n$) is
determined from the number of unique transmitters on the channel.

The measurement samples are updated every $T_{update}$.
In order to reduce the effect of short-term dynamics of traffic and channel
conditions, we maintain their exponential weighted average with 
smoothing parameter $\alpha$.
\begin{align*}
  \bar{T}_{tx} & \gets \alpha \bar{T}_{tx} + (1-\alpha) \hat{T}_{tx} \\
  \bar{R}_{tx} & \gets \alpha \bar{R}_{tx} + (1-\alpha) \hat{R}_{tx}
\end{align*}
Assuming that the new flow is from an inactive station,
\begin{equation*}
  n \gets n + 1
\end{equation*}
\subsection{Input flow specification}
The description of the traffic offered at the link-layer (referred 
to as $FlowSpec$) by a new flow will be 
provided by specifying the following parameters:
\begin{list}{$\bullet$} {\setlength{\itemsep}{0.01in}
  \setlength{\topsep}{0.02in} \setlength{\leftmargin}{0.25in}}
  \item packet arrival rate $\lambda_{flow}$
  \item average payload size in bits, $PAYLOAD_{flow}$
\end{list}
It should be noted that BUFFET makes use of the declared $FlowSpec$ only
while admitting that particular flow. For the previously admitted flows,
the aggregate flow statistics are derived from channel measurements and
thus inaccuracy as well as change in $FlowSpec$ will be automatically
adjusted before admitting the next flow.

\subsection{Deriving model inputs from measurements and $FlowSpec$}
For a moderately loaded WLAN in a steady state, 
all arrived packets at the interface queues are successfully transmitted 
on the channel. However, as
we are considering random packet arrival processes, momentary queue
buildup can happen when collisions occur. Therefore we approximate
packet arrival rate to the WLAN to be:
\begin{equation*}
  \lambda_{MAC} = R_{succ} + R_{coll} = \bar{R}_{tx} 
\end{equation*}

All admitted flows are able to obtain their required throughput that 
may be different from their fair share as long as the WLAN throughput
is less than the capacity.
As an approximation, the model assumes a
uniform $\lambda$ at every station. This approximation does not affect
the accuracy of the results as we are not using a station's fair share of throughput 
for admission decision. 
Thus, accounting for the new flow being admitted, $\lambda$ per station is then
averaged as
\begin{equation}
  \lambda_{new} = \frac{\lambda_{MAC}}{n} + \frac{\lambda_{flow}}{n}
\end{equation}

For a non-saturated WLAN, we ignore the effect of collision on the
measured frame duration. We
factor the payload and data rate for the new flow
by calculating $T_s^{flow}$, $T_s$ and $T_c$ as follows:
\begin{equation}
  \begin{split}
    T_s^{flow} &= DIFS + PHY\_HDR   \\ 
  	& \quad+ (MAC\_HDR+PAYLOAD_{flow})/R  \\
	   & \quad + SIFS + PHY\_HDR+ACK \\
	   T_s &= \frac{(\frac{\lambda_{MAC}}{n} \bar{T}_{tx} + \frac{\lambda_{flow}}{n} T_s^{flow})} {(\frac{\lambda_{MAC}}{n} + \frac{\lambda_{flow}}{n})} \\
  T_c &= T_s - (PHY\_HDR+ ACK + SIFS)
  \end{split}
\end{equation}
The $PHY\_HDR$ and $ACK$ in the above equations are expressed as
their respective durations and $R$ is the PHY data rate used by 
the station. 

\subsection{Admission Criterion}
\begin{figure}[htbp]
  \begin{center}
    \input{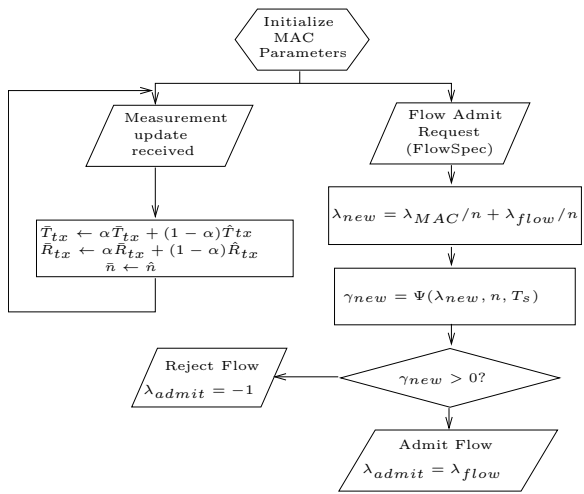}
  \end{center}
  \caption{Admission control flowchart}
  \label{fig:flowchart}
\end{figure}

As described earlier, a non-zero value of $\gamma$ indicates that the 
WLAN is not saturated. 
We use $\gamma_{new}$ predicted by the model as an indicator of saturation:
\begin{equation}
  \gamma_{new} = \Psi (\lambda_{new}, n, T_s)
\end{equation}
We admit a new flow only if the predicted value of $\gamma_{new}$ is non-zero. The BUFFET
algorithm is illustrated in Fig. \ref{fig:flowchart}.
%
%
\subsection{Description of other CAC mechanisms for performance comparison}
We compare BUFFET with centralized CAC mechanisms 
belonging to two other categories, namely saturation
model based CACs and threshold based CACs. 
Although they are originally proposed for EDCA, we adapt them to DCF
by considering only one access category as described next.

\subsubsection{Call Admission Control based on saturation model (TPUTSAT)}
According to the CAC mechanism 
proposed by Pong and Moors \cite{pong:call} 
based on Bianchi's saturation model \cite{bianchi:model} 
each station computes the probability $\tau$ of a transmission in a
slot from the measured probability $p$ of collision as
\begin{equation*}
  \tau = \frac{2(1-2p)}{(1-2p)(W+1) + p W(1- (2p)^m)}
\end{equation*}
from which $P_{tr}$, $P_{s}$ and $T_{slot}$ are obtained.
A flow is admitted if 
the achievable throughput by a station is sufficient to meet the
throughput demand:
\begin{align*}
  S_{flow} &= \frac{\tau (1-\tau)^{(n-1)} PAYLOAD_{flow} }{T_{slot}} \\
  \textrm{Admit if} & \quad  S_{flow} >= \textrm{Requested throughput} 
\end{align*}
The packet arrival rate $\lambda$ does not need to be supplied for
computation of $S_{flow}$ as it is the throughput at saturation.

\subsubsection{Threshold based admission control (AIRTIME)}
An airtime allocation and admission control is proposed 
in \cite{chou_achieving}.
Without consideration for parameterized QoS,
the airtime required per second by a flow $j$ from station $i$  
is 
\begin{equation*}
  r_{i,j} = \frac{s_{i,j}}{R_i} 
\end{equation*}
where $s_{i,j}$ is throughput requirement of flow $\{i,j\}$ and $R_i$ is
the PHY data rate used by station $i$.
Assuming the knowledge about all admitted flows, a new flow $q$ from station
$p$ is admitted if:
\begin{equation*}
  r_{p,q} + \sum_i \sum_j r_{i,j} \leq ~ EA
\end{equation*}
where $EA$ is the effective airtime ratio or airtime threshold that
excludes the control overhead of the resource allocation mechanism.
\section{Performance Analysis}
\label{perf}
We analyze the performance of BUFFET, TPUTSAT and AIRTIME through simulations
using the OPNET 11.5 modeler \cite{opnet}
%
according to the parameters given in Table \ref{tab:parameters}.
In each scenario, a new station requests a flow every 10 seconds.
All the flows have fixed payload size and Poisson packet arrivals
unless mentioned otherwise.
We compare the number of admitted flows (throughput)
and average
delay after the time when either admission control kicks in or saturation
sets in.

\begin{table}
  \centering
  \begin{tabular}{|c|c|}
    \hline
    Area & 50x50 m \\
    \hline
    Number of stations in WLAN & 10, 20, 40, 60 \\
    \hline
    Smoothing parameter $\alpha$ & 0.8 \\
    \hline
    Update interval $T_{update}$ & 1 sec. \\
    \hline
    Payload size & 100, 250, 500 and 1500 B \\
    \hline
    Packet arrival process & Poisson, CBR \\
    \hline
    PHY data rate & 2, 11 Mbps DSSS \\
    \hline
  \end{tabular}
  \caption{Simulation Parameters}
  \label{tab:parameters}
\end{table}

\begin{figure}[htbp]
  \begin{center}
    \includegraphics[width=3.0in]{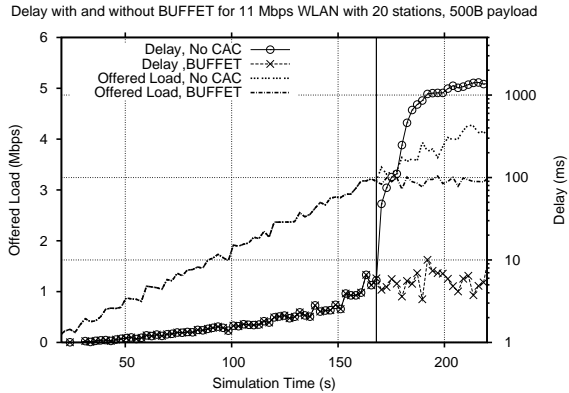}
  \end{center}
  \caption{Delays with and without CAC (BUFFET)}
  \label{fig:delaycompared}
\end{figure}

The delay vs. simulation time curves in Fig. \ref{fig:delaycompared}
with and without CAC (BUFFET) illustrate the working of the
CAC mechanism. At 170 seconds, BUFFET determines that the
requested flow would cause saturation and hence starts rejecting flows.
Accepting flows beyond this point causes the delay to rise sharply.

\begin{figure}[htbp]
    \includegraphics[width=2.6in,angle=270]{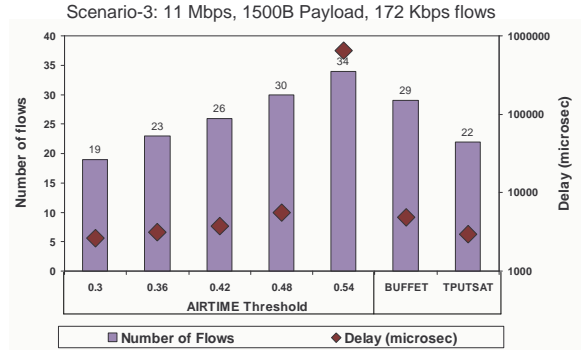}
  \caption{AIRTIME, BUFFET and TPUTSAT delay-throughput comparison}
  \label{fig:thresh_chart}
\end{figure}

Table \ref{tab:summ} summarizes the delay and throughput obtained with
BUFFET, TPUTSAT and AIRTIME for representative scenarios.
BUFFET is able to maintain the average delay under 7$ms$ for all scenarios. 
More importantly, this consistent delay performance is achieved
at a throughput close to the optimum.
For example, in scenario-1, BUFFET admits 27 flows; AIRTIME with a threshold
of 0.09 admits 30 flows but at the cost of WLAN saturation.
This aspect is pictorially depicted by Fig. \ref{fig:thresh_chart}
for scenario-3
which shows that BUFFET achieves high utilization at low delays,
managing a good balance between delay and utilization.

Table \ref{tab:summ} suggests that the delay and throughput for 
AIRTIME is a function of the airtime threshold. The optimum threshold
itself is variable across scenarios due to the effect of payload size 
and channel data rate on resource allocation overheads. 
Therefore, setting a correct threshold is essential for good performance
of AIRTIME.

On the other hand, both BUFFET and TPUTSAT avoid saturation and provide
low delays without depending on a threshold. TPUTSAT being based on
a saturation model provides marginally lower delays but conservatively
admits fewer number of flows. This effect is
more pronounced for higher $n$ when saturation models tend to
overestimate the effect of collision and contention. For instance,
for a WLAN size of 60 stations(scenario-5) TPUTSAT admits 40\% fewer flows
than BUFFET.

\begin{table*}
  \centering
  \subtable[Effect of payload size] {
  \begin{tabular}{|p{1.0in}||c|c|c|c|c|c|c|c|c|c|c|c|c|c|c|}
    \hline
    & \multicolumn{5}{c|}{scenario-1} & \multicolumn{5}{c|}{scenario-2} &
    \multicolumn{5}{c|}{scenario-3} \\
    \hline
    Sim. parameters& \multicolumn{5}{c|}{(11, 100, 32)} & \multicolumn{5}{c|}{(11, 500, 105)} &
    \multicolumn{5}{c|}{(11, 1500, 172)} \\
    \hline
    \multirow{2}{*}{CAC scheme} & \multirow{2}{*}{B} & \multirow{2}{*}{T} & \multicolumn{3}{c|}{Airtime} &
    \multirow{2}{*}{B} & \multirow{2}{*}{T} & \multicolumn{3}{c|}{Airtime} &
    \multirow{2}{*}{B} & \multirow{2}{*}{T} & \multicolumn{3}{c|}{Airtime} \\
    \cline{4-6} \cline{9-11} \cline{14-16}
    & & & 0.07 & 0.08 & 0.09 &
    & & 0.21 & 0.26 & 0.31 &
    & & 0.42 & 0.48 & 0.54 \\
    \hline
    Admitted flows & 27 & 23 & 24 & 27 & 30 &
    28 & 22 & 21 & 27 & 32 &
    29 & 22 & 26 & 30 & 34 \\
    \hline
    Delay ($ms$) & 4.36 & 1.71 & 1.90 & 4.36 & 4055 &
    3.41 & 1.94 & 1.82 & 2.92 & 1856 &
    4.82 & 2.99 & 3.76 & 5.55 & 652\\
    \hline
  \end{tabular}
  \label{tab:varpayload}
  }
  \\[0.1in]
  \subtable[Effect of data rate and flow bandwidth] {
  \begin{tabular}{|p{1.0in}||c|c|c|c|c|c|c|c|c|c|c|c|c|c|c|}
    \hline
    & \multicolumn{5}{c|}{scenario-4} & \multicolumn{5}{c|}{scenario-5} &
    \multicolumn{5}{c|}{scenario-6} \\
    \hline
    Sim. parameters& \multicolumn{5}{c|}{(2, 500, 33)} & \multicolumn{5}{c|}{(11, 500, 57)} &
    \multicolumn{5}{c|}{(11, 500, 400)} \\
    \hline
    \multirow{2}{*}{CAC scheme} & \multirow{2}{*}{B} & \multirow{2}{*}{T} & \multicolumn{3}{c|}{Airtime} &
    \multirow{2}{*}{B} & \multirow{2}{*}{T} & \multicolumn{3}{c|}{Airtime} &
    \multirow{2}{*}{B} & \multirow{2}{*}{T} & \multicolumn{3}{c|}{Airtime} \\
    \cline{4-6} \cline{9-11} \cline{14-16}
    & & & 0.47 & 0.54 & 0.61 &
    & & 0.23 & 0.26 & 0.29 &
    & & 0.26 & 0.31 & 0.36 \\
    \hline
    Admitted flows & 30 & 22 & 28 & 32 & 36 &
    50 & 31 & 44 & 50 & 55 &
    8 & 8 & 7 & 8 & 9\\
    \hline
    Delay ($ms$) & 6.52 & 4.24 & 5.62 & 8.23 & 2502 & 
    3.03 & 1.51 & 2.11 & 3.03 & 801 &
    5.83 & 5.83 & 3.06 & 5.83 & 412 \\
    \hline
  \end{tabular}
  \label{tab:varlambda}
  }
  \\[0.1in]
  \subtable[CBR flows] {
  \begin{tabular}{|p{1.0in}||c|c|c|c|c|c|}
    \hline
    & \multicolumn{5}{c|}{scenario-7} \\
    \hline
    Sim. parameters& \multicolumn{5}{c|}{(11, 500, 105)} \\
    \hline
    \multirow{2}{*}{CAC scheme} & \multirow{2}{*}{B} & \multirow{2}{*}{T} & \multicolumn{3}{c|}{Airtime} \\
    \cline{4-6}
    & & & 0.22 & 0.26 & 0.30 \\
    \hline
    Admitted flows & 24 & 26 & 22 & 27 & 31 \\
    \hline
    Delay ($ms$) & 1.36 & 1.68 & 1.29 & 1.87 & 1651\\
    \hline
  \end{tabular}
  \label{tab:cbr}
  }
  \begin{tabular}[t]{lp{2.2in}}
    \\
    Sim. parameters: & (PHY rate (Mbps), Payload size (B), Per-flow bandwidth) \\
    CAC scheme: & B: BUFFET, T: TPUTSAT\\ 
    & Airtime: AIRTIME with threshold \\
  \end{tabular}
  \caption{Admitted flows and delay comparison for representative scenarios}
  \label{tab:summ}
\end{table*}

For CBR flows (Table. \ref{tab:cbr}) BUFFET conservatively admits 
fewer flows than TPUTSAT owing to the assumption of Poisson packet arrival.
The loss of throughput is however marginal.
Lower measured probability of collision due to regular packet
arrivals helps TPUTSAT admit more flows. 

\begin{table}
  \centering
  \scriptsize{
  \begin{tabular}{|p{0.16in}|p{0.26in}|p{0.27in}||p{0.16in}|p{0.26in}|p{0.28in}||p{0.37in}|p{0.23in}|}
    \hline
    \multicolumn{3}{|c||}{Flow type-1} & \multicolumn{3}{c||}{Flow type-2} & &\\ 
    \hline
    PHY Rate & Payload (B) & Flow B/W (Kbps) & PHY Rate & Payload (B) & Flow B/W (Kbps) & Admitted flows & Delay ($ms$) \\
    \hline
    11 & 500 & 100 & 2 & 500 & 33 & 30 & 4.83 \\
    2 & 500 & 33 & 11 & 1500 & 172 & 30 & 6.76 \\
    11 & 100 & 32 & 11 & 1500 & 172 & 28 & 7.03 \\
    11 & 500 & 100 & 11 & 100 & 32 & 29 & 4.06 \\
    \hline
  \end{tabular} }
  \caption{BUFFET with for non-uniform flows}
  \label{tab:meamo}
\end{table}

As mentioned before, BUFFET adapts to non-uniform payload sizes as
well as data rates through $T_{tx}$ measurements. To verify this,
we conducted another set of simulations with BUFFET for two different types
of flows as listed in Table \ref{tab:meamo}. The first 20 flows
requested are of type-1 and next 20 flows are of type-2.
Delays in this case too are less than 7$ms$, confirming that BUFFET works well 
without any configuration even for combinations of diverse data rates and
application types.

BUFFET is therefore ideal for realistic WLAN deployments
with diverse applications and channel conditions, providing 
a fully distributed, zero-configuration autonomous setup.

\section{Conclusion}\label{conclusion}
In this work, we propose an autonomous distributed admission control
named BUFFET for 802.11 DCF that is based on an analytical model.
In order to keep the average delay within acceptable limits, BUFFET 
admits a flow only if it does not lead to WLAN saturation, an
indicator of which is a parameter $\gamma$ predicted by the model.
BUFFET is able to derive all inputs to the model from the measurements
(frame transmission rate, average transmission duration and number of 
stations) and requested FlowSpec.

Performance analysis through OPNET simulations suggests that
BUFFET is able to provide consistent sub-7$ms$ delay  
while achieving near-optimal throughput.
We also compare the performance of BUFFET with two other admission
control schemes, one based on saturation throughput (TPUTSAT) and the 
other based on airtime threshold (AIRTIME).
TPUTSAT is found to be too conservative in admitting flows, especially
for higher number of stations. Configuration of correct threshold
(which itself is widely variable based on load and data rate)
is essential for correct operation of AIRTIME.

The fully distributed nature of BUFFET, wherein it is able to
deduce information about already admitted flows, coupled with
its ability to work correctly for a combination of diverse data rates 
and payload sizes makes it ideal for zero-configuration 
self-regulating distributed WLAN setup.

We are currently implementing BUFFET for 
Atheros chipset based 802.11g WLAN cards on GNU/Linux systems.
Applying the algorithm to 802.11e EDCA by extending the model and 
using similar measurements per access category is another future
direction we are pursuing.

\bibliographystyle{abbrv}

\appendix  \label{our_model}
\section{DTMC model for non-saturated 802.11 DCF}

\begin{figure}
  \begin{center}
    \includegraphics[width=3.0in]{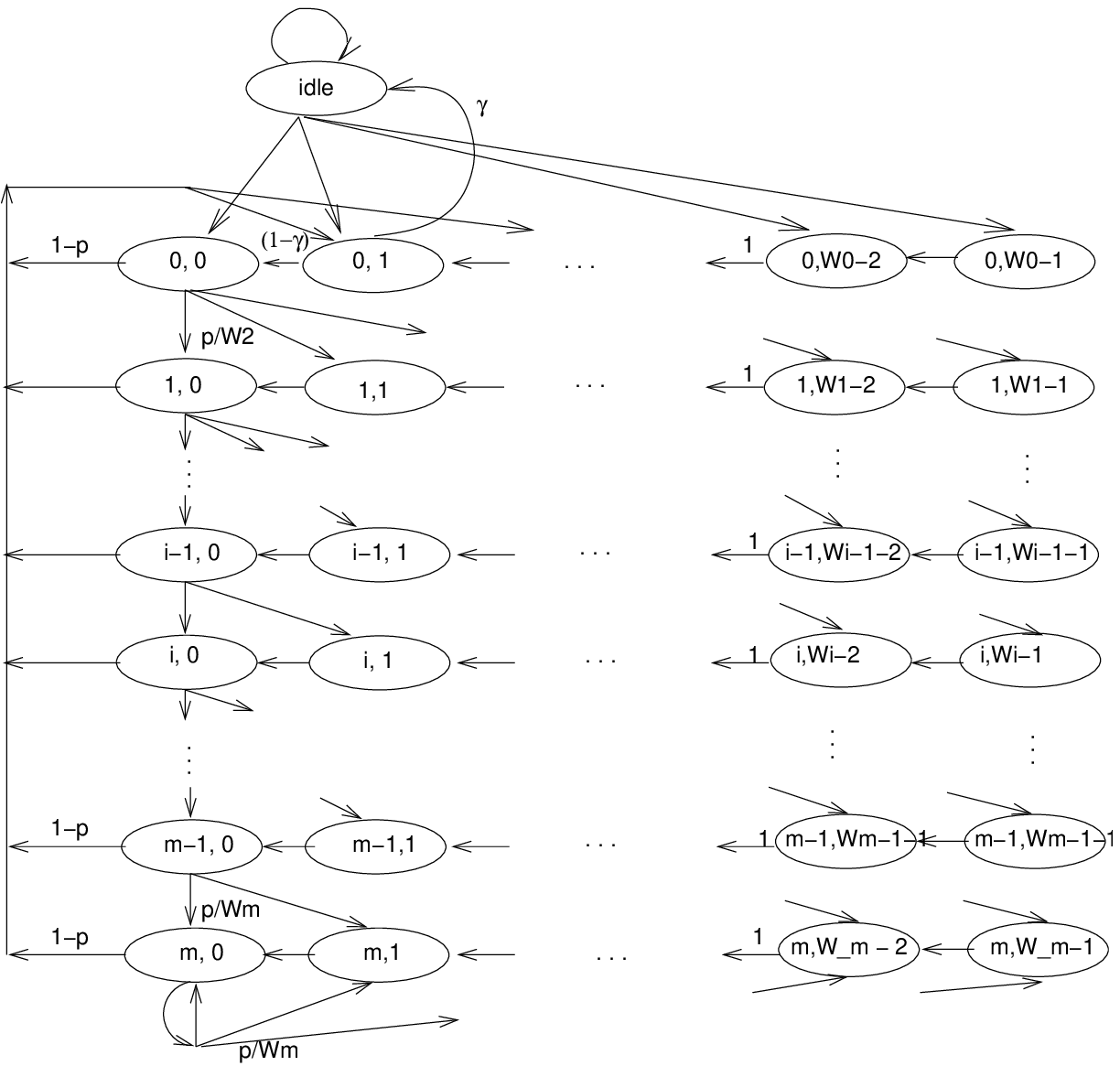}
  \end{center}
  \caption{Markov chain for a single station}
  \label{fig:chain}
\end{figure}

We model the behavior of a single station using a Discrete Time Markov 
Chain (DTMC). The model is based on the model proposed by Bianchi \cite{bianchi:model}
for saturated WLANs using a DTMC. 
To account for non-saturated conditions we add a state called \emph{idle} to this
DTMC (Fig. \ref{fig:chain}).
The station after completing a packet transmission performs a
mandatory backoff with random backoff counter picked from $(0,CW_{MIN}-1)$
\footnote{We follow the same terminology as that used in \cite{bianchi:model}.} 
(\emph{post-backoff}).
If the MAC transmission queue is empty after post-backoff, it goes in
\emph{idle} state. It remains in this state 
till the end of the slot corresponding to the first packet arrival.
We define the probability $\gamma$ as the probability that the station
queue is non-empty after the post-backoff.
At saturation, this probability becomes zero.

Accordingly, there are four possibilities after completion of post-backoff:
\begin{itemize}
  \item Station already has packets in its queue; it transmits back-to-back 
    packets after post-backoff without going in \emph{idle} state.
  \item A packet arrived in \emph{idle} state, and during a silent slot on the channel is
    transmitted in the next slot.
  \item A packet arrived in \emph{idle} state and during a slot corresponding to
    a transmission, before the transmission could be sensed (initial CCATime).
    The station is required to backoff in this case, similar to post-backoff.
  \item A packet arrived in \emph{idle} state and during a slot corresponding to a 
    (successful or collided) transmission but after the transmission has been sensed.
    The station is allowed to transmit after the end of the ongoing transmission 
    and the subsequent DIFS silence.
\end{itemize}

The state of the DTMC is defined by the tuple $\{s(t),b(t)\}$ as defined in
\cite{bianchi:model} or $idle$.
Similar to \cite{bianchi:model}, we assume a constant and independent conditional 
collision probability $p$
and ignore the relevant retry limits (ShortRetryLimit
or LongRetryLimit).
In addition, we assume that packet arrival to the MAC are Poisson with
rate $\lambda$.
We define $P_{as}, P_{ac}$, $P_{ai}$ and $P_a$ as the probabilities of 
a packet arrival
during successful transmission slot, collision slot, idle slot and a
generic slot respectively. 
\begin{equation} \begin{split}
  P_{as} &= 1- e^{-\lambda T_s} \\
  P_{ac} &= 1- e^{-\lambda T_c} \\
  P_{ai} &= 1- e^{-\lambda \sigma} \\
  P_{a} &= P_{tr}P_sP_{as} + P_{tr}(1-P_s)P_{ac} + (1-P_{tr})P_{ai}
\end{split} \end{equation}

The single-step transition probabilities are:
\begin{align*} \begin{split}
  P\{i,k ~|~ i,k+1\} &= 1, \quad k \in (0,W_i-2), ~ i\in (1,m) \\
  P\{0,k ~|~ i,0\} &= \frac{(1-p)}{W-1}, \quad  k \in (1,W-1), ~ i\in (0,m) \\
  P\{i,k ~|~ i-1,0\} &= \frac{p}{W_i}, \quad  k \in (0,W_i-1), ~ i\in (1,m) \\
  P\{m,k ~|~ m,0\} &= \frac{p}{W_m}, \quad  k \in (0,W_m-1) \\
  P\{idle ~|~ idle\} &= 1-P_a \\
  P\{0,k ~|~ idle\} &= \frac{P_{tr}(1-e^{-\lambda ~ CCA})}{W}, \quad k \in (1,W-1) \\
\end{split} \end{align*}
\begin{align*} \begin{split}
  P\{0,0 ~|~ idle\} &= P_{tr}P_s(e^{-\lambda ~ CCA} - e^{-\lambda ~ T_s}) \\
  & \quad + P_{tr}(1-P_s)(e^{-\lambda ~ CCA} - e^{-\lambda ~ T_c}) \\
  & \quad + (1-P_{tr})P_{ai} + \frac{P_{tr}(1-e^{-\lambda ~ CCA})}{W} \\
  P\{idle ~|~ 0,1) &= \gamma \\
  P\{0,0 ~|~ 0,1\} &=	1-\gamma \\
  P\{0,k ~|~ 0,k+1\} &= 1, \quad  k \in (1,W-1)
\end{split} \end{align*}

This is an aperiodic, irreducible Markov chain for which steady state probabilities
are known to exist. 
Applying the normalization condition
\begin{equation}
  1 = \sum_{i=0}^m \sum_{k=0}^{W_i} b_{i,k} + b_{idle} 
\end{equation}
and solving the chain,
the probabilities $\tau$, $P_{tr}$, $P_s$ and throughput $S$ are obtained similar to 
\cite{bianchi:model}.


\subsection{Calculation of MAC service time ($D_{MAC}$)}
The MAC service time is the duration from the time a 
packet becomes head of the queue to the time it is acknowledged. 
The number of transmissions before the packet is successfully transmitted
follows a modified geometric distribution with parameter $p$.
The delays and probabilities for the four possibilities after 
successful transmission mentioned above are calculated as follows:
\begin{align*}
  D_{b2b} &= T_s + \frac{T_c ~ p}{1-p} + BO_{slots} \\
  P_{b2b} &= 1 - \gamma \\
  D_{nob2b\_idle} &= D_{b2b} - \frac{W ~ T_{slot}}{2} \\
  P_{nob2b\_idle} &= \frac{\gamma (1-P_{tr})P_{ai}}{P_a} \\
  D_{nob2b\_tx} &= D_{b2b} - \frac{W ~ T_{slot}}{2} + \frac{[P_s ~ T_s + (1-P_s) ~ T_c]}{2} \\
  P_{nob2b\_tx} &= \frac{ \gamma [ P_{tr} P_s (e^{- \lambda ~ CCA} - e^{- \lambda ~ T_s}) ]}{P_a} \\ 
  & \quad + \frac{\gamma [P_{tr} (1 - P_s) (e^{- \lambda ~ CCA} - e^{- \lambda ~ T_c}) ]} {P_a} \\
  D_{nob2b\_cca} &= D_{b2b} + P_s ~ T_s + (1-P_s) ~ T_c \\
  P_{nob2b\_cca} &= \frac{ \gamma P_{tr} (1 - e^{- \lambda ~ CCA})}{P_a}
\end{align*}
where 
  $BO_{slots} = \frac{(1-p)W ~ T_{slot}}{2} \left[ \frac{2(1-2p^m)}{1-2p} + 
  \frac{[(2p)^m(2-p) - (1-p) ]}{(1-p)^2} \right] $  
  and $T_{slot}$ is the average duration of a logical slot.
The average MAC service time, $D_{MAC}$, is the conditional average of the above.

\subsection{Calculation of $\gamma$}
Treating the WLAN interface as an $M/G/1$ queue,
from the definition of queue utilization ($\rho$) and $\gamma$, we obtain 
\begin{align} \label{eqn:gamma} \begin{split}
  \rho &= \textrm{min}(1, \lambda/\mu) = \textrm{min}(1, \lambda ~ D_{MAC}) \\
  \gamma &= P\{\text{MAC queue is empty}\} = 1 - \rho 
\end{split} \end{align}
The values of $\gamma$, $\tau$ and $D_{MAC}$ are obtained numerically 
through successive iterations. For convenience, we express $\gamma$ as a function
of three load-dependent variables:
\begin{equation}
  \gamma = \Psi(\lambda, n, T_s)
\end{equation}



\end{document}